\documentclass{ifacconf}
\usepackage{amsmath}
\usepackage{graphicx}      
\usepackage{natbib}        
\usepackage{subcaption}
\usepackage{graphicx}
\usepackage{tikz}
\usepackage{comment}
\usepackage[]{todonotes}

\begin{document}
\begin{frontmatter}
\title{Health-Aware Fast Charging Using Homogenized Model with Heterogeneous Internal State Reconstruction}

\author[First,Fourth]{Alessio A. Lodge,}
\author[Third]{Alessio Lombardo Pontillo,}
\author[First]{Feye S.J. Hoekstra,}
\author[First]{Robinson Medina,}
\author[First,Fifth]{Steven Wilkins,}
\author[Fourth]{Ilenia Battiato}
\address[First]{TNO, Powertrains Department, Helmond, The Netherlands}
\address[Fourth]{Stanford University, California, USA}
\address[Third]{DISAT Department, Politecnico di Torino, Turin, Italy}
\address[Fifth]{Eindhoven University of Technology, Eindhoven, The Netherlands}
\thanks{\textcopyright\ 2026 the authors. This work has been accepted to IFAC for publication under a Creative Commons Licence CC-BY-NC-ND.}

\begin{abstract}                
Fast charging of lithium-ion batteries is limited by lithium plating, which occurs when the anode potential drops below 0 V vs Li/Li+. Model-based control aims to maximize charging current while maintaining anode potentials above this threshold. In this work, a plating-free fast charging strategy is demonstrated using a Homogenized Model (HM) coupled with a classical PID controller. The HM, derived from homogenization theory applied to the Poisson-Nernst-Planck equations, retains the physics of the Doyle-Fuller-Newman model while capturing electrode microstructural heterogeneity in a one-dimensional double-continua formulation. By reconstructing three-dimensional distributions of electrochemical variables from precomputed closure variables, the HM enables non-invasive estimation of heterogeneous anode potentials, acting as a virtual sensor. Through MATLAB–COMSOL co-simulation, a PID controller regulates current to maintain the full 3D anode potential distribution above the plating limit, achieving model-based fast charging at a fraction of the computational cost of high-fidelity models. The results demonstrate the potential of HM-based control for safe, degradation-aware, and efficient fast charging of lithium-ion batteries.
\end{abstract}

\end{frontmatter}

\section{Introduction}

The electrification of transport, powered by renewable energy, is an effective method to reduce global greenhouse gas emissions \citep{SACCHI2022112475}. The adoption of electric vehicles (EVs) can be accelerated by improvements in charging time \citep{corradi_what_2023}. Advancements in battery materials, engineering, and improved control can increase charging speeds and shift refueling time closer to that of internal combustion engine vehicles \citep{an2020key}. A battery can be charged much faster than the nominal rate, however, at the expense of lifespan due to increased degradation \citep{waldmann2018li}. This occurs due to a dominant side reaction in fast charging called lithium plating. With high charging currents, at the surface of an active material, lithium will preferentially undergo metallic deposition, rather than intercalating \citep{gao2021interplay}. This phenomenon is a source of loss of performance and degradation of cell characteristics \citep{ahmed2017enabling}.

Modifying operating conditions, for the same cell and overall charge time, may lead to reduced degradation \citep{mathieu2021comparison}. As such, implementing smart charging protocols can enable fast charging \citep{epding2020aging, koleti2019development}. It has been shown experimentally that maintaining anode potential, measured with a reference electrode, above 0 V vs Li/Li+, reduces degradation induced during charging, as it avoids lithium plating, which occurs for anode potentials equal to or below 0 V vs Li/Li+ \citep{Rangarajan2020, adam2020fast}. 
Since commercial lithium-ion cells are not equipped with reference electrodes and such an addition is invasive, the anode potential cannot be directly measured, rather it may be estimated. Model-based control of fast charging aims at enforcing the charging conditions that minimize degradation and maximize charging speed, based on the information, such as the estimated anode potential, provided by a model \citep{Wassiliadis2023}. Using a model to estimate anode potential constitutes a virtual sensor.

 Electrochemical models (also called physics-based models - PBMs) are battery models that are employed to simulate the behavior  of lithium-ion batteries \citep{brosa2022continuum} and investigate the evolution in time and space of physical states such as concentrations and potentials. The Doyle-Fuller-Newman model (DFN), developed by \cite{doyle1993modeling}, is a well-known model that considers an idealized battery system, consisting of three domains, the negative electrode, the separator and the positive electrode and two phases, the liquid phase (electrolyte) and solid phase (particle active material). It is a pseudo-2D reduced-order model based on coupling of micro- and macro-scale dynamics, where electrolyte concentration and potentials are described in an average sense across the battery thickness ($x$-direction), and remain coupled to the local solution of a microscale problem for the concentration and potential in the electrode particle, assumed to be spherical at each macroscopic location $x$. The DFN remains a go-to physics-based modeling tool compared to, e.g., microscale models. Yet, the advancement in computational capabilities has made more sophisticated multiscale models accessible, including models based on homogenization theory \citep{Battiato2029Theory}, such as the model derived in \cite{arunachalam2015veracity}, here referred to as the Homogenized Model.

 The Homogenized Model (HM) is based on homogenization theory by multiple scale expansion and presents a double-continua macroscale formulation where, unlike the micro-macro formulation in the DFN, the electrolyte and the electrode domains are described as interpenetrating continua governed by coupled partial differential equations describing the spatiotemporal evolution of the macro/average potential and lithium-ion concentration in the solid and electrolyte domains. The HM has additional important features that distinguish it from the more commonly used DFN. Firstly, it does not rely on the assumption that the electrode is composed of  spherical particles, and instead allows for realistic electrode complexity in the direct calculation of its parameters \citep{Weber2022Homogenization}. Secondly, it allows for the rigorous reconstruction of the microscale/porescale internal states from the macroscopic quantities through a closure variable \citep{Weber2022Homogenization}. This is critical to capture battery scale performance degradation due to localized microscopic effects at a fraction of the computational cost compared to solving full microscale models.  By reconstructing the 3D distribution of model states, the HM preserves the 3D nature of electrochemical variables and enables more precise control applications, such as fast charging algorithms. Parameterization efforts require similar approaches as for the DFN, with the addition of geometrical information obtained via FIB/SEM or CT measurements   \citep{Weber2022Homogenization}.

In this work, a HM-based fast charging framework is demonstrated, achieved via PID feedback control. The major contribution is the use of the HM-simulated 3D distribution of the anode potential as the physical constraint for the fast charging strategy. The model and controller run at a fraction of the computational cost of microscale models, low enough to be potentially suitable for embedded applications, maximizing current while keeping the entire 3D distribution of anode potential above the plating threshold, superseding DFN-like approaches.

\section{Homogenized Model}


The majority of electrochemical models used in literature for model-based fast charging are based on the DFN model \citep{andersson2024electrochemical, kolluri2020real, Robinson}, reduced-order versions such as the SPM(e) \citep{pramanik2016electrochemical, Wassiliadis2023}, other bespoke reductions \citep{chu2017non, pathak2017analyzing}, and hybrid ML-DFN approaches \citep{lin2019real}. The reason for its popularity is that the DFN strikes an important balance between computational effort and accuracy, compared to other models like microscale models \citep{schmidt2021understanding}, and hence has been demonstrated to be suitable for battery management system (BMS) deployment and use on embedded systems \citep{oehler2022embedded}. 

\begin{figure}[h]
    \centering
    \includegraphics[width=0.95\linewidth]{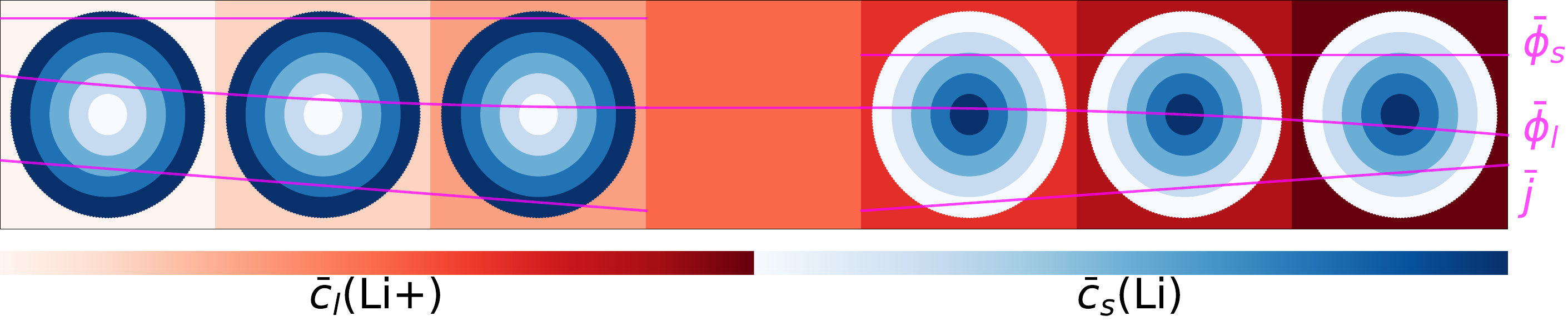}
    \vspace{0.5em}
    \includegraphics[width=0.95\linewidth]{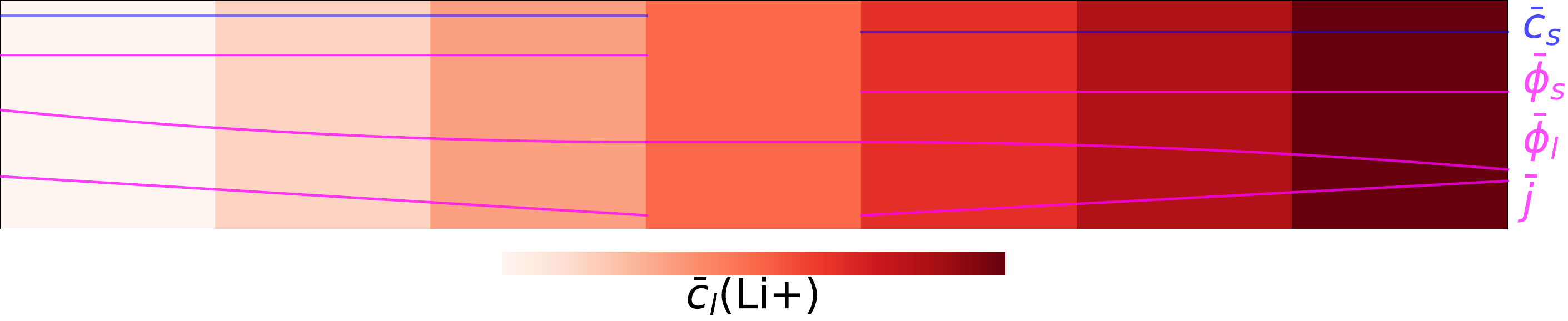}
    \caption{DFN (top) and HM (bottom). Domains: pseudo-2D and 1D double-continua. States: $c^s$, $c^\ell$, $\phi^s$, $\phi^\ell$.}
    \label{fig:DFN_HM}
\end{figure}

The DFN estimates concentrations and potentials in the solid phase (across the $x$-direction and for each $x$, also in the radial dimension $r$) and in the liquid phase (solely in the $x$ dimension) \citep{marquis2020suite}. The resulting anode potential $\phi^-$ (the difference between solid potential at the particle surface and electrolyte potential at the same position) therefore is unique, for each location $x$, as spherical symmetry is assumed. In reality, electrodes are three-dimensional highly heterogeneous structures \citep{xu2021guiding}, far from spherical, hence at each $x$ position, there is a large scatter in magnitude of the in-plane anode potentials and all other spatially resolved states \citep{latz2015multiscale}. Hence, it is proposed to use the HM, rather than the DFN, as it leverages the heterogeneous microstructural information and is capable of capturing spatial heterogeneity, while maintaining low computational cost for model-based control. The two modeling frameworks are briefly summarized in Figure \ref{fig:DFN_HM}. 

\begin{figure}[h]
  \centering
  \includegraphics[width=0.48\columnwidth]{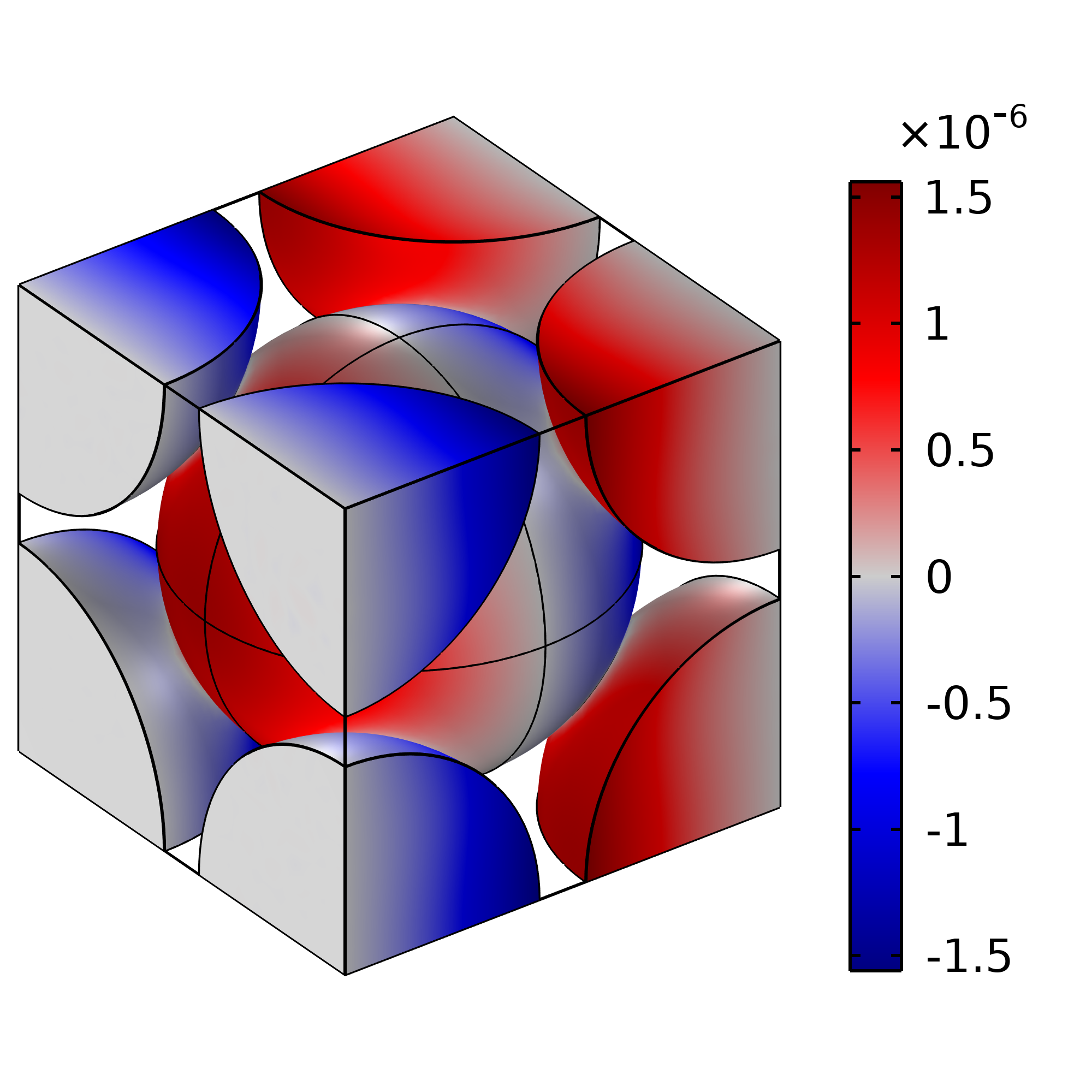}
  \hfill
  \includegraphics[width=0.48\columnwidth]
  {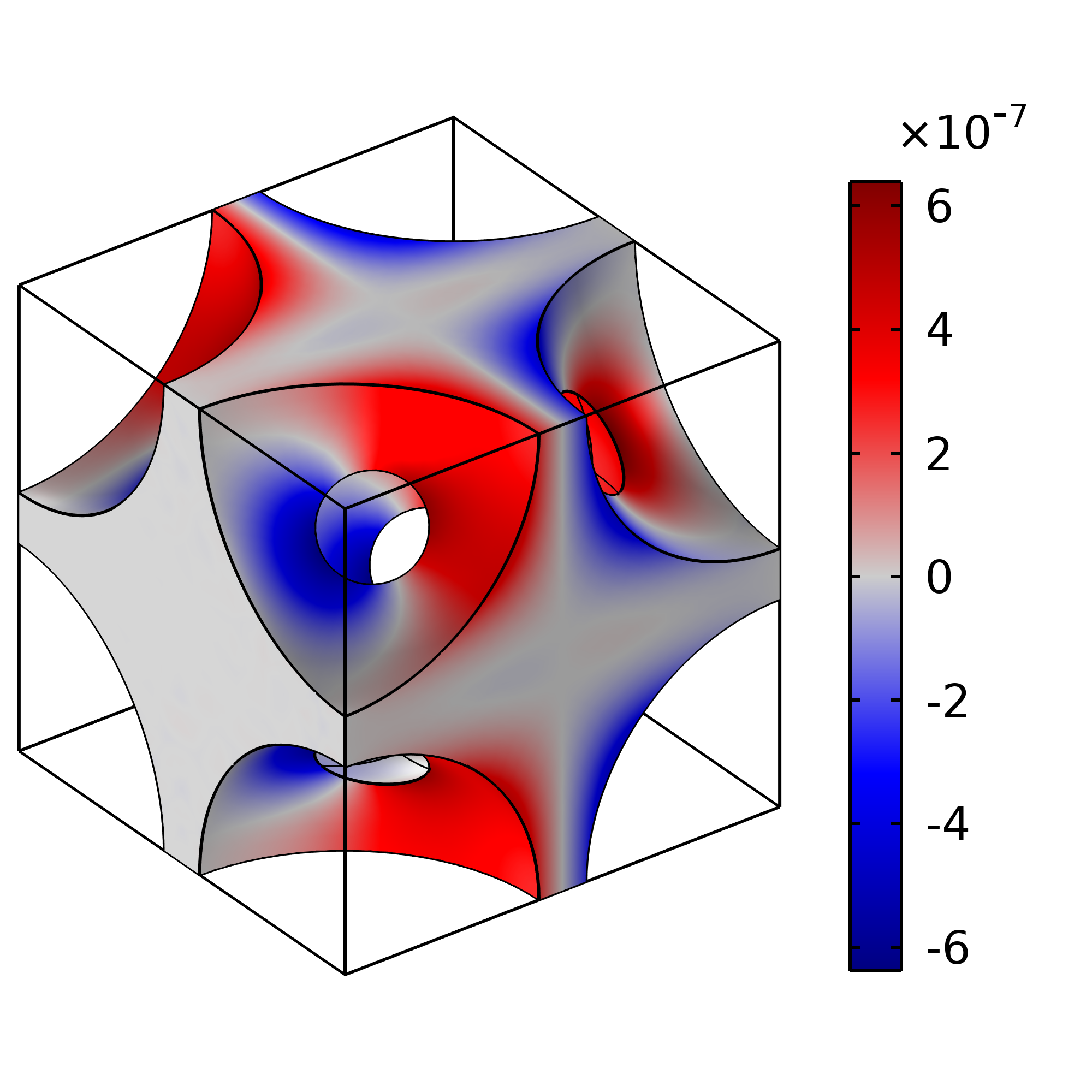}
  \caption{Closure variables $\chi^s$ (left, solid phase) and  $\chi^\ell$ (right, liquid phase) from the solution of the BVP in Equations (\ref{eq:BVP1},\,\ref{eq:BVP2}).} 
  \label{fig:unit_cell_problem}
\end{figure}

The HM has its theoretical foundation in the Poisson-Nernst-Planck (PNP) equations, the same utilized to describe porous electrode theory common to the DFN model \citep{arunachalam2015veracity}. By employing homogenization theory \citep{hornung1997homogenization}, the PNP equations are upscaled to a system of PDEs resolved on a 1D (2D or 3D) double-continua, divided into three domains: negative electrode, separator and positive electrode. The model equations, derived in \cite{arunachalam2015veracity}, are as follows, in which Equations \eqref{eq:HM_mass_liquid} and \eqref{eq:HM_mass_solid} express the conservation of mass of lithium-ions and lithium in the liquid phase and solid phase, respectively, \eqref{eq:HM_charge_liquid} and \eqref{eq:HM_charge_solid} express the conservation of charge in the liquid phase and solid phase, respectively, and \eqref{eq:HM_BV} expresses the interfacial current as a Butler-Volmer relationship.
\begin{equation}
\begin{split}
\eta \frac{\partial c^\ell}{\partial t} 
+ \nabla_\mathbf{x}\cdot \Biggl[\Bigl(
-\mathbf{D}^{\ell}_{eff} 
+ \frac{2RT t_{+}(1 - t_{+})}{F^2} 
\mathbf{K}^{\ell}_{eff} \frac{1}{c^\ell} \Bigr) \nabla_\mathbf{x}c^\ell \\
- \frac{t_{+}}{F} \mathbf{K}^{\ell}_{eff} \nabla_\mathbf{x}\phi^\ell 
\Biggr] 
= 2 \eta  \varepsilon^{-1}\frac{\mathcal{K}^\star}{F} i_{\text{ct}}
\end{split}
\label{eq:HM_mass_liquid} 
\end{equation}
\begin{equation}
\frac{\partial c^s}{\partial t}
- \nabla_\mathbf{x}\cdot\Bigl( \mathbf{D}^{s}_{eff} \nabla_\mathbf{x}c^s \Bigr)
= -2 \varepsilon^{-1}  \eta  \frac{\mathcal{K}^\star}{F}  i_{\text{ct}}
\label{eq:HM_mass_solid}
\end{equation}
\begin{equation}
\begin{split}
\nabla_\mathbf{x}\cdot \Biggl[
\Bigl( -\frac{2RT(1 - t_{+})}{F} \mathbf{K}^{\ell}_{eff}\frac{1}{c^\ell} \Bigr) \nabla_\mathbf{x}c^\ell
+ \mathbf{K}^{\ell}_{eff} \nabla_\mathbf{x}\phi^\ell
\Biggr] \\
=  -2 \eta \varepsilon^{-1} \mathcal{K}^\star i_{\text{ct}}
\label{eq:HM_charge_liquid}
\end{split}
\end{equation}
\begin{equation}
\nabla_\mathbf{x}\cdot \Biggl[ \mathbf{K}^{s}_{eff} \nabla_\mathbf{x}\phi^s \Biggr]
= 2  \varepsilon^{-1}  \eta  \mathcal{K}^\star i_{\text{ct}}
\label{eq:HM_charge_solid}
\end{equation}
\begin{equation}
i_{\text{ct}} = k\,2 \sqrt{c^s c^\ell} \sqrt{1 - \frac{c^s}{c^s_\text{max}}}
\sinh\!\left( \frac{F U}{2RT} \right)
\label{eq:HM_BV}
\end{equation}
where $c^\ell,c^s$ are the average liquid and solid concentrations; $\phi^\ell,\phi^s$ the average potentials; $\mathbf{D}^{\ell}_{eff},\mathbf{D}^{s}_{eff}$ effective diffusivities; $\mathbf{K}^{\ell}_{eff},\mathbf{K}^{s}_{eff}$ effective conductivities; $t_+$ transference number; $F$ Faraday constant; $R$ ideal gas constant; $T$ temperature; $\varepsilon$ the ratio between macro and micro length scales, also called separation variable; $\eta$ porosity; $\mathcal{K}^\star$ volumetric interfacial area; $i_{\mathrm{ct}}$ interfacial current; $k$ rate constant; $c^s_\text{max}$ max concentration; $U$ overpotential. 
The effective parameters, such as diffusivity and conductivity, are in general tensorial, and are defined as follows:
\begin{equation}
\mathbf{A}^{j}_{eff}=\frac{1}{V^j}\int_{V^j}\mathbf{A}^{j}(\nabla_\mathbf{y} \chi^j + \mathbf{I})\,\text{d}\mathbf{y} 
\end{equation}
where $j=\{\ell,s\}$, $\mathbf{A}=\{\mathbf{D},\mathbf{K}\}$,  $V^j$ is the volume of the corresponding phase and $\chi^j$, also referred to as closure variables, are defined as the solutions to the following boundary value problem (BVP) on the unit cell \citep{korneev2020data}:
\begin{subequations}
\begin{equation}
    \nabla_\mathbf{y} \cdot (\nabla_\mathbf{y} \chi^j + \mathbf{I}) = 0, 
    \quad \mathbf{y} \in \mathcal{J}, 
    \label{eq:BVP1}
\end{equation}
\begin{equation}
    \mathbf{n}_j \cdot (\nabla_\mathbf{y} \chi^j + \mathbf{I}) = 0, 
    \quad \mathbf{y} \in \Gamma.
    \label{eq:BVP2}
\end{equation}
\end{subequations}
with $\mathcal{J} =\{\mathcal{B},\mathcal{S}\}$, where $\mathcal{B}$ and $\mathcal{S}$ are the liquid and solid phase domains, and $\Gamma$ is the liquid-solid domain interface. The solution to the BVP, i.e. the closure variables $\chi^j$, are shown in Figure \ref{fig:unit_cell_problem}.

The stereological information of the electrode microstructures is conserved within the effective transport parameters through $\chi^j$. In the case of an isotropic microstructure, the effective parameters reduce to  $\mathbf{A}^{j}_{eff}=\beta^j\mathbf{I}$ and the model becomes effectively 1D. Since no assumption is made on the microscale, other than periodicity of the unit cell, the HM can accommodate highly heterogeneous microstructures \citep[e.g.,][]{Weber2022Homogenization, korneev2020data}. This differs from the DFN approach, which in principle assumes a Bruggeman relationship for the effective parameters, with $A_{eff} = \beta A$ and $\beta = \eta^\alpha$, with $\alpha$ the Bruggeman coefficient, equal to 1.5 and does not allow for internal state reconstruction or microstructure informed calculation of effective parameters.

\section{3D Reconstruction}

\begin{figure}[h!]
  \vspace{-1.5cm}
  \hspace{-1cm}
  \centering
  \begin{tikzpicture}
    \def\overlap{1cm}
    \hspace{-0.5cm}
    \node[anchor=south west, inner sep=0] at (0,-\overlap)
      {\includegraphics[width=1.3\linewidth]{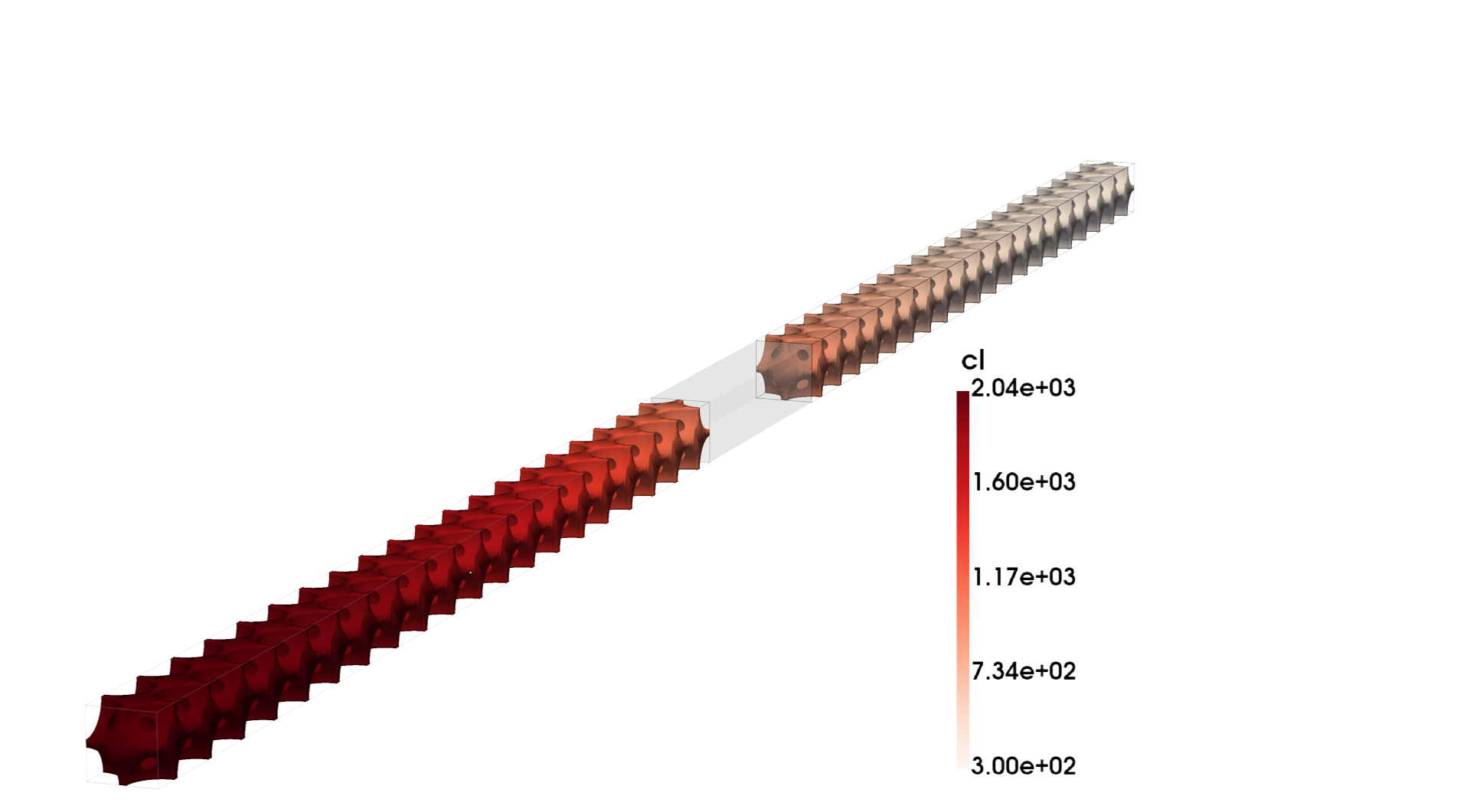}};
    \node[anchor=south west, inner sep=0] at (0,0)
    {\includegraphics[width=1.3\linewidth]{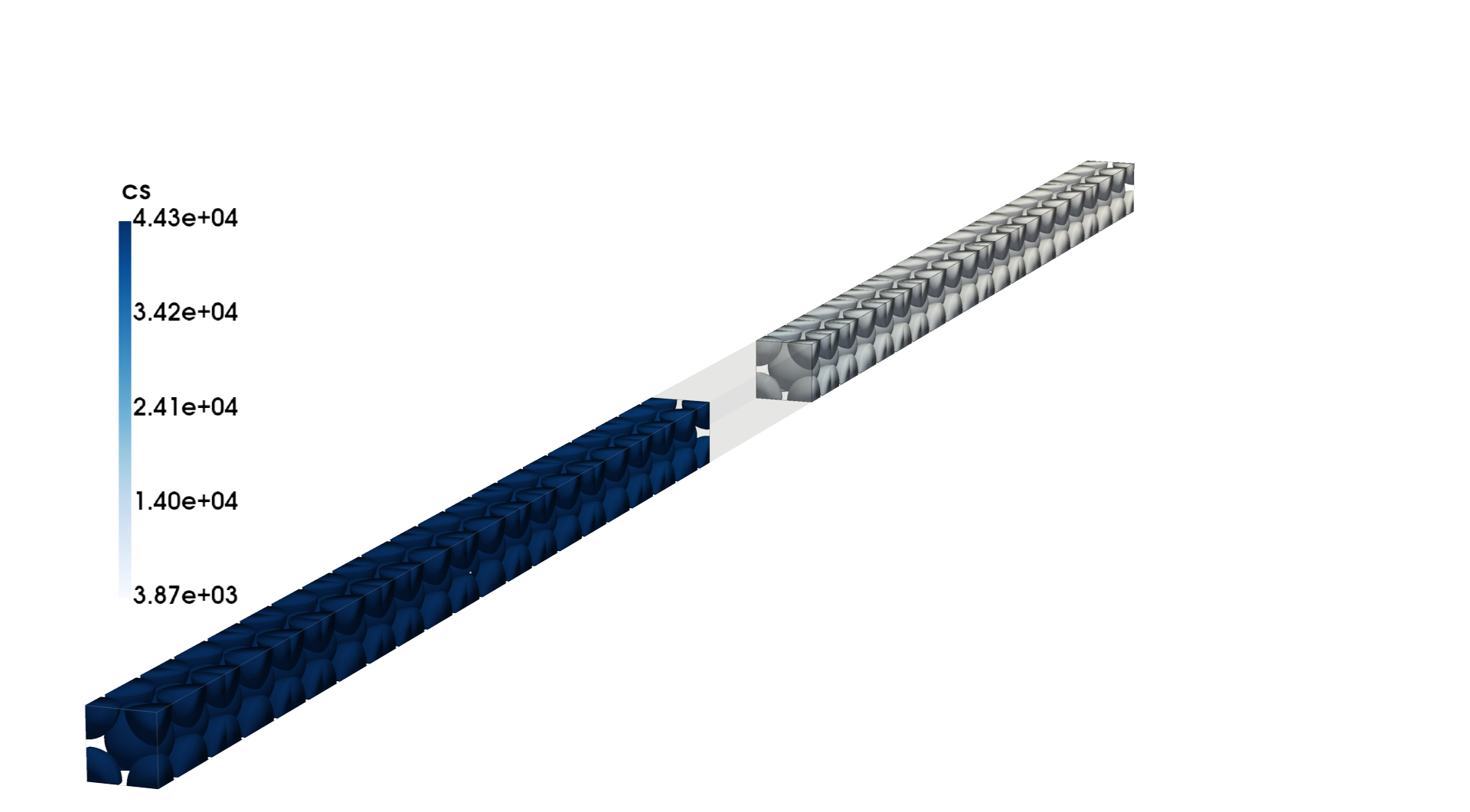}};
  \end{tikzpicture}
  \caption{Snapshot of the 3D reconstruction of $c^s$ and $c^\ell$ distributions, from Equation \eqref{eq:recasting}, HM solutions and closure variables $\chi(x,y,z)$.}
  \label{fig:3D_Reconstruction}
\end{figure}

It can be shown (see \cite{Battiato2029Theory}) that the internal states for the concentration and potential in the liquid and solid phases can be approximated as a truncated series in the form of
\begin{equation}
\Phi(x,y,z) = \Phi_0 + \varepsilon\, \chi(x,y,z)\, \nabla \Phi_0 + \mathcal{O}(\varepsilon^2)
\label{eq:recasting}
\end{equation}
in which $\Phi$ is a reconstructed internal state, $\Phi_0$ (e.g. $c^s$, $c^\ell$, $\phi^s$ or  $\phi^\ell$) is the solution of the HM, $\varepsilon$ is the separation variable and $\chi^j$ is the respective closure variable. In other words, such ``unresolved" states can be calculated from the average quantities ($c^s$, $c^\ell$, $\phi^s$ or  $\phi^\ell$), solution of (\ref{eq:HM_mass_liquid}--\ref{eq:HM_BV}), and the precalculated $\chi^j$ from the BVP (\ref{eq:BVP1},\,\ref{eq:BVP2}). This implies that the solution of the HM can then be mapped in 3D, through the closure variables $\chi^j$ to obtain the heterogeneous distribution of each electrochemical variable \citep{Battiato2029Theory}. In Figure \ref{fig:3D_Reconstruction}, a snapshot of the reconstructed surface concentrations of lithium in the solid phase $c^s$ and lithium-ions in the liquid phase $c^\ell$ is shown, both obtained through reconstruction as in Equation \eqref{eq:recasting}.

For demonstration purposes and without loss of generality, the battery parameters and microstructure are taken from the microscale model provided in \cite{lombardo2025comparative}, in which the transport, kinetic and thermodynamic parameters are those of a lithium-ion cell (Graphite$\vert$NMC111), whilst the microstructure is idealized as an orderly packing of spherical particles. For the application of homogenization theory, the microstructure is periodic, with the full electrodes constituted by the repetition of a periodic unit cell. Scale separation is ensured by constructing both electrodes from 20 periodic Face-Centered Cubic (FCC) unit cells, and quantified by $\varepsilon$ equal to $0.05$ $(1/20)$ for this geometry. Through the solution of the BVP, the values of $\beta^j$ are equal to 0.19650 for the liquid and 0.22375 for the solid phase. It is emphasized that arbitrarily complex microstructures can be accommodated in this framework as long as they are spatially 3D periodic, as shown in \cite{Weber2022Homogenization}.

\section{Fast Charging Methodology}
A charging strategy is, at minimum, the time series of current applied to a battery from an initial battery state $x_0$ to a final state $x_f$, typically based on state-of-charge (SOC) (i.e. $SOC_0$ to $SOC_f$). Additional operating conditions could also be prescribed, such as temperature or pressure. Leveraging feedback control for fast charging allows to utilize the information on the system, based on a model, to construct the charging profile.

\begin{figure}[h!]
    \centering
    \includegraphics[width=1.0\linewidth]{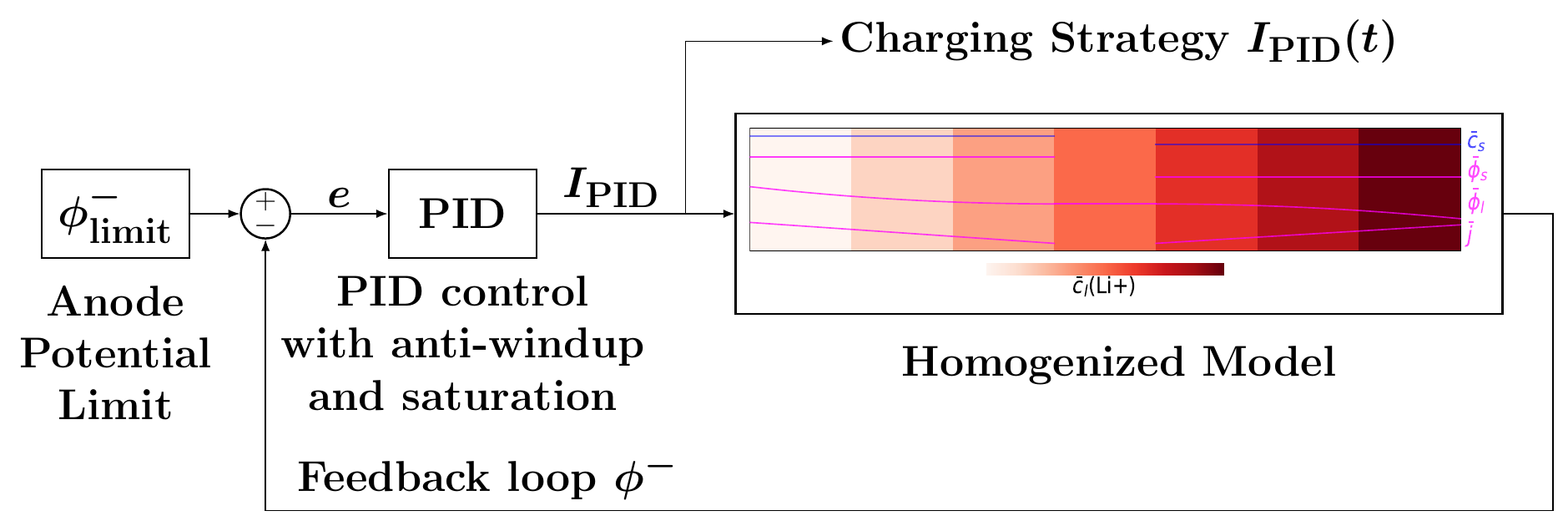}
    \caption{HM-based fast charging with PID controller.}
    \label{fig:PID}
\end{figure}

To demonstrate the use of a HM with 3D reconstruction of electrochemical variables for fast charging and compare it to existing work, which applies a DFN model (or simplified versions of it), a similar control strategy to that applied in \citep{Wassiliadis2023} is adopted. Here, a feedback controller (PID) is used to regulate the current so that the estimated anode potential $\phi^-$ is as low as possible, i.e., close to, but above, the plating threshold. Controlling the anode potential in this way naturally leads to maximum current, i.e., fastest charging. The control loop applied in this paper is visualized in Figure \ref{fig:PID} and includes a saturation limit and anti-windup. Other safety conditions, such as terminal voltage and cell temperature control (the model used here is isothermal), are to be considered in follow-up work. Formally, the applied control current is given by
\begin{equation}
I_{\text{PID}}(t) = \max(0, \ \min(\ I_{\phi^-}(t), I_{\text{max}})),\label{eq:PIDCurrent}
\end{equation}
with $I_{\text{max}}$ being the maximum control current, $I_{\text{PID}}(t)$ the saturated applied current, and $I_{\phi^-}(t)$ the unsaturated PID output given by
\begin{equation}
I_{\phi^-}(t) = K_P\, e(t) + K_I \int_{0}^{t} e(\tau)\, d\tau + K_D \frac{d e(t)}{dt},
\end{equation}
with $e(t) = \phi^-_{}(t) - \phi^-_{\text{limit}}(t)$, with $\phi^-_{\text{limit}}(t)$ being the plating threshold including buffer (in this case 10 mV, see \cite{frank2024investigating} for discussions on the buffer) and $\phi^-(t)$ the estimated anode potential. In the case of a DFN-type model, the applied $\phi^-(t)$ is the idealized single value anode potential occurring at the particle/electrolyte interface closest to the separator. In contrast, for the HM with 3D reconstruction, the estimated anode potential is reconstructed spatially in three dimensions: first in the through-thickness 1D direction of the electrode, by solving the underlying HM PDEs, and then mapped via reconstruction to 3D. In this case, $\phi^-$ is chosen as the minimum value of the 3D distribution, i.e. $\phi^-(t) = \min(\phi^-(t,x,y,z))$. This forces the entire `band' of anode potentials above the threshold as seen in Figure \ref{fig:HM_3D_vs_avg_bands}. In practice, as the 1D anode potential, estimated by the HM, is monotonic in space during charge, the lowest 3D anode potential occurs in the periodic unit cell closest to the separator, hence only the last unit cell of the negative electrode is reconstructed to identify $\phi^-(t)$. The values of $K_P$, $K_I$ and $K_D$, obtained via manual tuning, are 2.0, 0.0005, and 0.5, respectively.

Simulations are performed using a combination of MATLAB, for the controller, and COMSOL, for solving the electrochemical model. COMSOL LiveLink is leveraged to enable continuous communication between the two software suites, in a co-simulation setup.  An initial current of $I(0) = I_{\text{max}}$ is used. The simulation runs from 0\% SOC until an end condition of 80\% SOC is reached. Fixed time stepping of 1 s is also implemented, with numerical tolerance of 1E-4 in COMSOL and a \textit{fine} mesh scheme with 9 uniformly distributed discretization points in each electrode. The framework was run on a portable laptop, an HP EliteBook 860, with 32 GB RAM and an Intel i7 12th Generation 1.8 GHz processor. Although the approach is demonstrated in simulation, it sets the foundation for future exploration with embedded deployment and battery-in-the-loop (BIL) validation. Indeed, the HM does not present the nested dependency of the DFN between the micro- and macro-scale, in which a solver must iteratively compute the solution on both scales at each time step. Hence, in this formulation, the HM is computationally lighter, suggesting the opportunity for embedded deployment, as demonstrated for the DFN in \cite{oehler2022embedded}.

\section{Results}

\begin{figure}[h!]
  \centering
  \includegraphics[width=\linewidth]{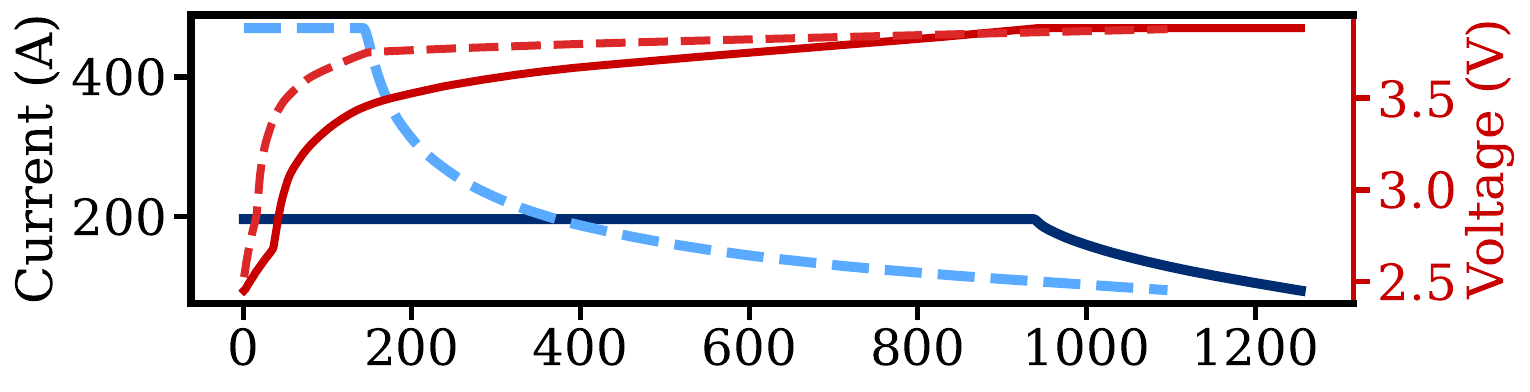}
    \includegraphics[width=\linewidth]{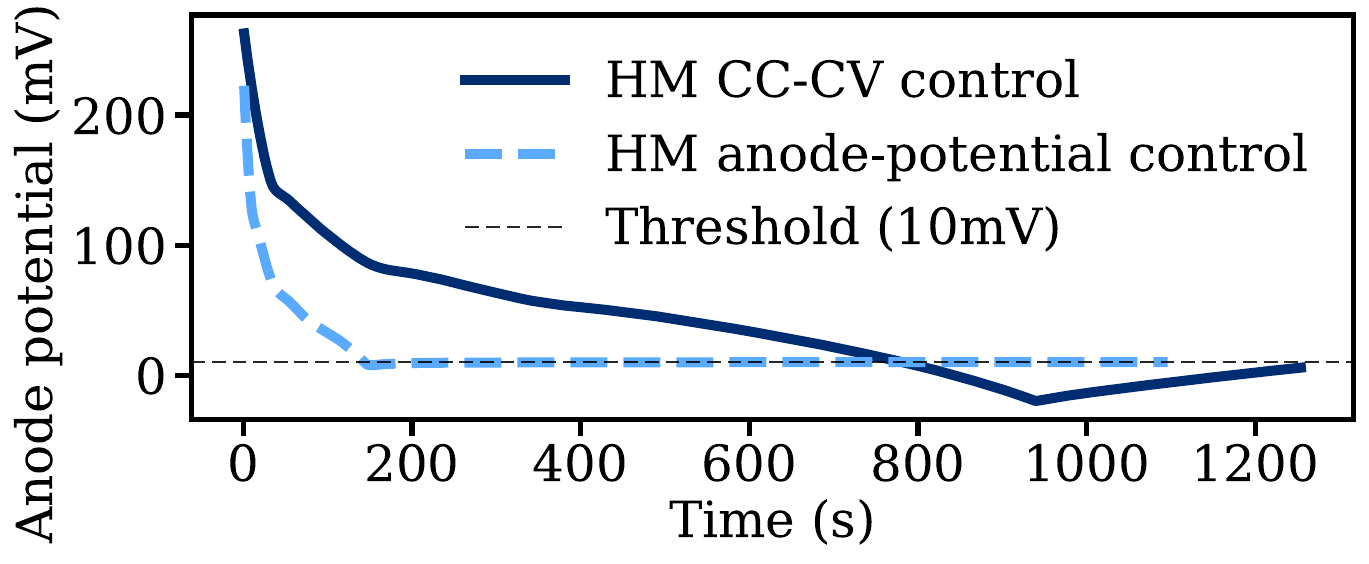}
    
  \vspace{-0.5em}
  \caption{CC-CV vs \ HM anode potential control.}
  \label{fig:CC-CV_vs_anode_potential_control}
\end{figure}

\begin{figure}[h!]
  \vspace{-0.5em}
  \centering
  \includegraphics[width=\linewidth]{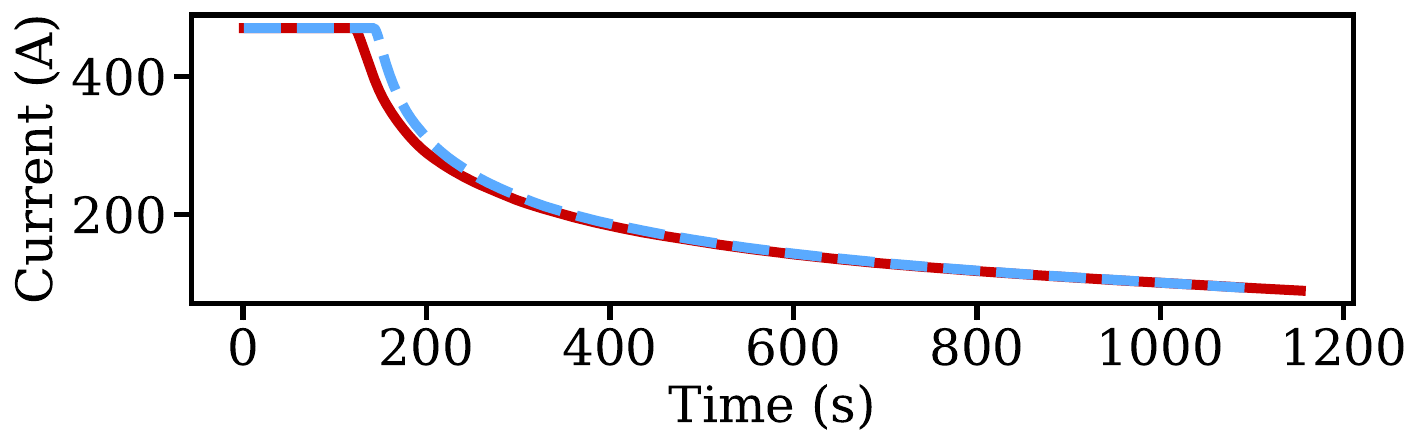}
  \vspace{0.01em}
  \includegraphics[width=\linewidth]{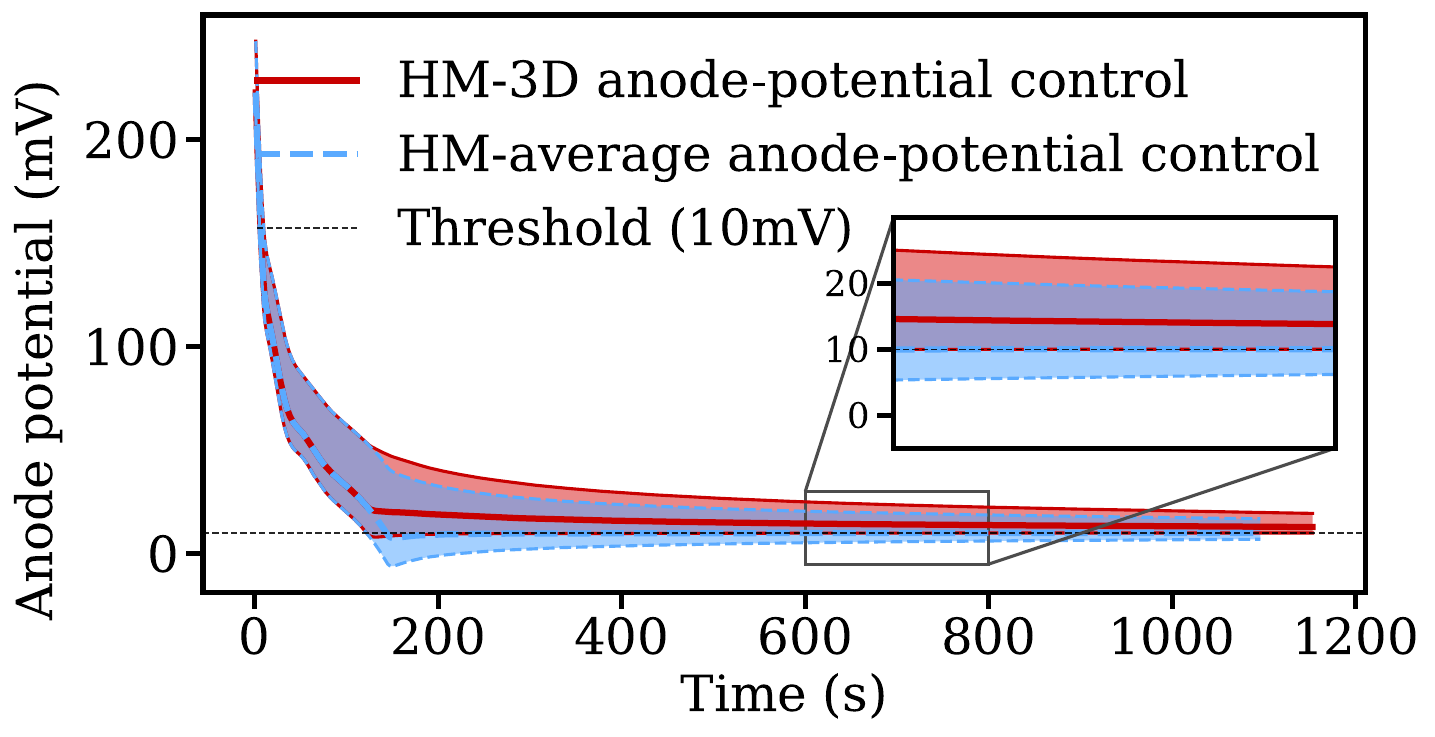}
  \vspace{-1.6em}
  \caption{HM-3D anode potential control vs \ HM-average anode potential control. Band displays reconstructed 3D distribution of interfacial anode potential.}
  \label{fig:HM_3D_vs_avg_bands}
\end{figure}

Model-based fast charge is successfully implemented by utilizing a HM in a classical PID feedback control framework, run jointly between COMSOL and MATLAB. For the battery cell chosen, i.e. for the selected parameters and microstructure, a charging time of 18.3 minutes was achieved, a 12.6\% reduction when compared to the constant-current constant-voltage (CC-CV) charging experiment (with an assumed datasheet-prescribed $I_{\text{max}}$ of 2.5 C used in the CC-CV, compared with the 6 C $I_{\text{max}}$ permitted in the electrochemical PID controller). Figure \ref{fig:CC-CV_vs_anode_potential_control} displays the response of the cell for the CC-CV, which leads to an extensive period of time (37.8\% of total) with negative anode potentials. In contrast, by operating above the manufacturer-specified limits on current and enforcing an electrochemical model-based control that limits anode potential, the proposed methodology achieves an increase in charging speed and a reduction in degradation, simultaneously. Depending on the cell design, properties, and consequently the parameters, such increase in charging speeds may be even larger and the time spent at negative anode potentials during CC-CV may be more severe.
It is possible to avoid not only the HM average anode potentials breaching the lithium plating threshold, but also any local anode potentials of the 3D heterogeneous distribution. This can be achieved in a reduced-order fashion by using the HM and the 3D reconstruction (see Equation \ref{eq:recasting}). For this, a distinction is made between HM-3D and HM-average, the former a controller that uses the HM output with 3D reconstruction and the latter that uses the HM output without 3D reconstruction. Naturally, as can be observed in Figure \ref{fig:HM_3D_vs_avg_bands}, the resulting HM-3D anode potential control charging strategy (HM-3DAPCS) is more conservative. In the HM-average anode potential control charging strategy (HM-APCS), despite the average anode potential remaining above the threshold, there are locally anode potentials below the plating threshold. Although the HM-3DAPCS has slightly smaller currents, fast charging is achieved, only 4.9\% slower than the HM-APCS and still 8\% faster than the CC-CV.

Interestingly, the average anode potential from HM-3DAPCS (thick middle line in the lower plot in Figure \ref{fig:HM_3D_vs_avg_bands}) looks skewed towards the lower limit of the 3D distribution band, despite the model having a symmetrical microstructure, with symmetrical closure variables and bands. This is due to the average anode potential of Figure \ref{fig:HM_3D_vs_avg_bands} being the HM solution output evaluated at the extremity of the negative electrode. However, for the projection in 3D and the calculation of the scatter band, the HM outputs $\phi^s$ and $\phi^\ell$ are utilized, evaluated at the center of the last unit cell, positioned half a unit cell distance before the extremity (rather than at the extremity). Thus, because of the monotonicity of the HM output anode potential, the band is calculated from a value slightly higher than that of the extremity.
The effectiveness of the protocol, which should be validated experimentally, depends on the accuracy of the anode potential prediction, which in turn relies on operating within the model’s applicability conditions and on parameterization. The DFN and HM can at best replicate high-fidelity microscale models, provided their applicability conditions are satisfied, but even then, issues of imprecise parameterization and unidentifiability remain \citep{plett2024perspectives}.

\section{Conclusion}
With the aim of reducing charging time to improve EV adoption, model-based fast charging is demonstrated by leveraging conventional control methods - a PID with anti-windup and a saturation limit - in a novel conjunction with an advanced electrochemical model, the Homogenized Model (HM), developed by \cite{arunachalam2015veracity}. The HM is shown to be a suitable plant model of a battery cell for health-aware control, with the additional feature of capturing microstructure-level electrode heterogeneity. 
By estimating the 3D distribution of the relevant internal states within the electrode, the control algorithm can avoid not only degradation-triggering average conditions (i.e. in this case anode potentials $\leq$ 10 mV, with 0 mV the plating threshold and 10 mV the  buffer), but also avoid any of these conditions occurring locally. Hence, by utilizing the PID to maintain the entire 3D distribution of anode potentials above the lithium plating threshold, health-aware fast charging can be achieved. This method improves on CC-CV charging protocols, not only in charge time but also by avoiding lithium plating operating conditions entirely.
In this study, CC-CV would lead to a charging time of 21 minutes, with 37.8\% of the time spent below the plating threshold. With the HM, without 3D reconstruction, the charging time reduces to 18.3 minutes (a 12.6\% improvement) with, on average, no time spent below the plating threshold. When investigating, via HM with 3D reconstruction, the same charging strategy, it is shown that this leads to significant portions of the charging protocol with anode potentials locally below the threshold - despite the average anode potential being above threshold. Only the approach that estimates the 3D distribution of anode potentials enables fast charge (19.2 min total, an 8\% reduction with respect to CC-CV) without any local plating risk.
\bibliography{ifacconf} 

@article{Weber2022Homogenization,
title = {Homogenization-Informed Convolutional Neural Networks for Estimation of Li-ion Battery Effective Properties},
journal = {Transport in Porous Media},
volume = {145},
number = {2},
pages = {527--548},
year = {2022},
author = {Weber, Ross M. and Korneev, Svyatoslav and Battiato, Ilenia},
}

@article{Battiato2029Theory,
  title={Theory and applications of macroscale models in porous media},
  author={Battiato, Ilenia and Ferrero V, Peter T and O’Malley, Daniel and Miller, Cass T and Takhar, Pawan S and Valdes-Parada, Francisco J and Wood, Brian D},
  journal={Transport in Porous Media},
  volume={130},
  number={1},
  pages={5--76},
  year={2019},
  publisher={Springer}
}

@article{SACCHI2022112475,
title = {When, where and how can the electrification of passenger cars reduce greenhouse gas emissions?},
journal = {Renewable and Sustainable Energy Reviews},
volume = {162},
pages = {112475},
year = {2022},
issn = {1364-0321},
author = {R. Sacchi and C. Bauer and B. Cox and C. Mutel},
}

@article{an2020key,
  title={Key strategies to increase the rate capacity of cathode materials for high power lithium-ion batteries},
  author={An, Qi and Sun, Xiaohong and Guo, Jinze and Cai, Shu and Zheng, Chunming},
  journal={Journal of the Electrochemical Society},
  volume={167},
  number={14},
  pages={140528},
  year={2020},
  publisher={IOP Publishing}
}

@article{gao2021interplay,
  title={Interplay of lithium intercalation and plating on a single graphite particle},
  author={Gao, Tao and Han, Yu and Fraggedakis, Dimitrios and Das, Supratim and Zhou, Tingtao and Yeh, Che-Ning and Xu, Shengming and Chueh, William C and Li, Ju and Bazant, Martin Z},
  journal={Joule},
  volume={5},
  number={2},
  pages={393--414},
  year={2021},
  publisher={Elsevier}
}

@article{ahmed2017enabling,
  title={Enabling fast charging--A battery technology gap assessment},
  author={Ahmed, Shabbir and Bloom, Ira and Jansen, Andrew N and Tanim, Tanvir and Dufek, Eric J and Pesaran, Ahmad and Burnham, Andrew and Carlson, Richard B and Dias, Fernando and Hardy, Keith and others},
  journal={Journal of Power Sources},
  volume={367},
  pages={250--262},
  year={2017},
  publisher={Elsevier}
}

@article{waldmann2018li,
  title={Li plating as unwanted side reaction in commercial Li-ion cells--A review},
  author={Waldmann, Thomas and Hogg, Bj{\"o}rn-Ingo and Wohlfahrt-Mehrens, Margret},
  journal={Journal of Power Sources},
  volume={384},
  pages={107--124},
  year={2018},
  publisher={Elsevier}
}

@article{mathieu2021comparison,
  title={Comparison of the impact of fast charging on the cycle life of three lithium-ion cells under several parameters of charge protocol and temperatures},
  author={Mathieu, Romain and Briat, Olivier and Gyan, Philippe and Vinassa, Jean-Michel},
  journal={Applied energy},
  volume={283},
  pages={116344},
  year={2021},
  publisher={Elsevier}
}

@article{epding2020aging,
  title={Aging-Optimized Fast Charging of Lithium Ion Cells Based on Three-Electrode Cell Measurements},
  author={Epding, Bernd and Rumberg, Bj{\"o}rn and Mense, Maximilian and Jahnke, Hannes and Kwade, Arno},
  journal={Energy Technology},
  volume={8},
  number={10},
  pages={2000457},
  year={2020},
  publisher={Wiley Online Library}
}

@article{arunachalam2015veracity,
  title={On veracity of macroscopic lithium-ion battery models},
  author={Arunachalam, Harikesh and Onori, Simona and Battiato, Ilenia},
  journal={Journal of The Electrochemical Society},
  volume={162},
  number={10},
  pages={A1940},
  year={2015},
  publisher={IOP Publishing}
}

@article{brosa2022continuum,
  title={A continuum of physics-based lithium-ion battery models reviewed},
  author={Brosa Planella, Ferran and Ai, Weilong and Boyce, Adam M and Ghosh, Abir and Korotkin, Ivan and Sahu, Smita and Sulzer, Valentin and Timms, Robert and Tranter, Thomas G and Zyskin, Maxim and others},
  journal={Progress in Energy},
  volume={4},
  number={4},
  pages={042003},
  year={2022},
  publisher={IOP Publishing}
}

@article{Rangarajan2020,
  title={Anode potential controlled charging prevents lithium plating},
  author={Rangarajan, Sobana P and Barsukov, Yevgen and Mukherjee, Partha P},
  journal={Journal of Materials Chemistry A},
  volume={8},
  number={26},
  pages={13077--13085},
  year={2020},
  publisher={Royal Society of Chemistry}
}

@article{Wassiliadis2023,
  author  = {Wassiliadis, N. and Kriegler, J. and Abo Gamra, K. and Lienkamp, M.},
  title   = {Model-based health-aware fast charging to mitigate the risk of lithium plating and prolong the cycle life of lithium-ion batteries in electric vehicles},
  journal = {J. Power Sources},
  volume  = {561},
  pages   = {232586},
  year    = {2023},
}

@article{corradi_what_2023,
	title = {What drives electric vehicle adoption? {Insights} from a systematic review on {European} transport actors and behaviours},
	volume = {95},
	issn = {22146296},
	shorttitle = {What drives electric vehicle adoption?},
	language = {en},
	urldate = {2025-09-22},
	journal = {Energy Research \& Social Science},
	author = {Corradi, Chiara and Sica, Edgardo and Morone, Piergiuseppe},
	month = jan,
	year = {2023},
	pages = {102908},
}

@article{frank2024investigating,
  title={Investigating anode potential errors of real-time capable dfn type models induced by inhomogeneity for fast charging of cylindrical lithium-ion batteries},
  author={Frank, Alexander and Durdel, Axel and Scheller, Maximilian and Sturm, Johannes and Jossen, Andreas},
  journal={Journal of The Electrochemical Society},
  volume={171},
  number={7},
  pages={070520},
  year={2024},
  publisher={IOP Publishing}
}

@article{oehler2022embedded,
  title={Embedded real-time state observer implementation for lithium-ion cells using an electrochemical model and extended Kalman filter},
  author={Oehler, Fabian F and N{\"u}rnberger, Kajetan and Sturm, Johannes and Jossen, Andreas},
  journal={Journal of Power Sources},
  volume={525},
  pages={231018},
  year={2022},
  publisher={Elsevier}
}

@article{schmidt2021understanding,
  title={Understanding deviations between spatially resolved and homogenized cathode models of lithium-ion batteries},
  author={Schmidt, Adrian and Ramani, Elvedin and Carraro, Thomas and Joos, Jochen and Weber, Andr{\'e} and Kamlah, Marc and Ivers-Tiff{\'e}e, Ellen},
  journal={Energy Technology},
  volume={9},
  number={6},
  pages={2000881},
  year={2021},
  publisher={Wiley Online Library}
}

@article{pramanik2016electrochemical,
  title={Electrochemical model based charge optimization for lithium-ion batteries},
  author={Pramanik, Sourav and Anwar, Sohel},
  journal={Journal of Power Sources},
  volume={313},
  pages={164--177},
  year={2016},
  publisher={Elsevier}
}

@article{adam2020fast,
  title={Fast-charging of automotive lithium-ion cells: In-situ lithium-plating detection and comparison of different cell designs},
  author={Adam, A and Wandt, J and Knobbe, E and Bauer, G and Kwade, A},
  journal={Journal of The Electrochemical Society},
  volume={167},
  number={13},
  pages={130503},
  year={2020},
  publisher={IOP Publishing}
}

@article{koleti2019development,
  title={The development of optimal charging strategies for lithium-ion batteries to prevent the onset of lithium plating at low ambient temperatures},
  author={Koleti, Upender Rao and Zhang, Cheng and Malik, Romeo and Dinh, Truong Quang and Marco, James},
  journal={Journal of Energy Storage},
  volume={24},
  pages={100798},
  year={2019},
  publisher={Elsevier}
}

@article{chu2017non,
  title={Non-destructive fast charging algorithm of lithium-ion batteries based on the control-oriented electrochemical model},
  author={Chu, Zhengyu and Feng, Xuning and Lu, Languang and Li, Jianqiu and Han, Xuebing and Ouyang, Minggao},
  journal={Applied energy},
  volume={204},
  pages={1240--1250},
  year={2017},
  publisher={Elsevier}
}

@article{andersson2024electrochemical,
  title={Electrochemical model-based aging-adaptive fast charging of automotive lithium-ion cells},
  author={Andersson, Malin and Streb, Moritz and Prathimala, Venu Gopal and Siddiqui, Aamer and Lodge, Andrew and Klass, Verena L{\"o}fqvist and Klett, Matilda and Johansson, Mikael and Lindbergh, G{\"o}ran},
  journal={Applied Energy},
  volume={372},
  pages={123644},
  year={2024},
  publisher={Elsevier}
}

@article{pathak2017analyzing,
  title={Analyzing and minimizing capacity fade through optimal model-based control-theory and experimental validation},
  author={Pathak, Manan and Sonawane, Dayaram and Santhanagopalan, Shriram and Braatz, Richard D and Subramanian, Venkat R},
  journal={ECS transactions},
  volume={75},
  number={23},
  pages={51},
  year={2017},
  publisher={IOP Publishing}
}

@article{kolluri2020real,
  title={Real-time nonlinear model predictive control (NMPC) strategies using physics-based models for advanced lithium-ion battery management system (BMS)},
  author={Kolluri, Suryanarayana and Aduru, Sai Varun and Pathak, Manan and Braatz, Richard D and Subramanian, Venkat R},
  journal={Journal of The Electrochemical Society},
  volume={167},
  number={6},
  pages={063505},
  year={2020},
  publisher={IOP Publishing}
}

@INPROCEEDINGS{Robinson,
  author={Medina, Róbinson and Hoedemaekers, Erik and Wilkins, Steven},
  booktitle={Proc. IEEE VPPC}, 
  title={Health-Conscious Charging of {L}i-ion Battery Cells: Using {P}{B}{M}s to Minimize Calendar and Cyclic Ageing Effects}, 
  year={2023},
}

@article{lin2019real,
  title={Real-time prediction of anode potential in li-ion batteries using long short-term neural networks for lithium plating prevention},
  author={Lin, Xianke},
  journal={Journal of The Electrochemical Society},
  volume={166},
  number={10},
  pages={A1893},
  year={2019},
  publisher={IOP Publishing}
}

@article{marquis2020suite,
  title={A suite of reduced-order models of a single-layer lithium-ion pouch cell},
  author={Marquis, Scott G and Timms, Robert and Sulzer, Valentin and Please, Colin P and Chapman, S Jon},
  journal={Journal of The Electrochemical Society},
  volume={167},
  number={14},
  pages={140513},
  year={2020},
  publisher={IOP Publishing}
}

@article{lombardo2025comparative,
  title={Comparative Analysis via CFD Simulation on the Impact of Graphite Anode Morphologies on the Discharge of a Lithium-Ion Battery},
  author={Lombardo Pontillo, Alessio and Marcato, Agnese and Versaci, Daniele and Marchisio, Daniele and Boccardo, Gianluca},
  journal={Batteries},
  volume={11},
  number={7},
  pages={252},
  year={2025},
  publisher={MDPI}
}

@inproceedings{plett2024perspectives,
  title={Perspectives on Methods to Overcome Obstacles to Physics-Based Models of LIB for EV},
  author={Plett, Gregory L and Trimboli, M Scott},
  booktitle={2024 International Conference on Electrical, Computer and Energy Technologies},
  pages={1--6},
  year={2024},
  organization={IEEE}
}

@article{xu2021guiding,
  title={Guiding the design of heterogeneous electrode microstructures for Li-ion batteries: microscopic imaging, predictive modeling, and machine learning},
  author={Xu, Hongyi and Zhu, Juner and Finegan, Donal P and Zhao, Hongbo and Lu, Xuekun and Li, Wei and Hoffman, Nathaniel and Bertei, Antonio and Shearing, Paul and Bazant, Martin Z},
  journal={Advanced Energy Materials},
  volume={11},
  number={19},
  pages={2003908},
  year={2021},
  publisher={Wiley Online Library}
}

@book{hornung1997homogenization,
  title={Homogenization and porous media},
  author={Hornung, Ulrich},
  volume={6},
  year={1997},
  publisher={Springer Science \& Business Media}
}

@article{korneev2020data,
  title={A data-driven multiscale framework to estimate effective properties of lithium-ion batteries from microstructure images},
  author={Korneev, Svyatoslav and Arunachalam, Harikesh and Onori, Simona and Battiato, Ilenia},
  journal={Transport in Porous Media},
  volume={134},
  number={1},
  pages={173--194},
  year={2020},
  publisher={Springer}
}

@article{latz2015multiscale,
  title={Multiscale modeling of lithium ion batteries: thermal aspects},
  author={Latz, Arnulf and Zausch, Jochen},
  journal={Beilstein journal of nanotechnology},
  volume={6},
  number={1},
  pages={987--1007},
  year={2015},
  publisher={Beilstein-Institut}
}

@article{doyle1993modeling,
  title={Modeling of galvanostatic charge and discharge of the lithium/polymer/insertion cell},
  author={Doyle, Marc and Fuller, Thomas F and Newman, John},
  journal={Journal of the Electrochemical society},
  volume={140},
  number={6},
  pages={1526--1533},
  year={1993},
  publisher={The Electrochemical Society, Inc.}
}

\end{document}